\documentclass[sigconf]{acmart}
\usepackage[utf8]{inputenc}
\usepackage{xspace}
\usepackage{subcaption}
\usepackage{hyperref}
\usepackage{tikz,graphicx}

\newcommand{\dashboard}{\textsc{TA-Dash}\xspace}
\newcommand{\au}{\textit{affected}(u,t)\xspace}

\title[\dashboard]{\dashboard: An Interactive Dashboard for Spatial-Temporal Traffic Analytics - Demo Paper}

\author{Nicolas Tempelmeier$^{1}$, Anzumana Sander$^{2}$, Udo Feuerhake$^{3}$, Martin Löhdefink$^{2}$, Elena Demidova$^{1}$}
\affiliation{%
  \institution{$^{1}$L3S Research Center, Leibniz Universit\"at Hannover, Hannover, Germany}
   \institution{$^{2}$PROJEKTIONISTEN GmbH, Hannover, Germany}
   \institution{$^{3}$Institute of Cartography and Geoinformatics, Leibniz Universit\"at Hannover, Hannover, Germany}
}
\email{{tempelmeier, demidova}@L3S.de, {sander, loehdefink}@projektionisten.de, Udo.Feuerhake@ikg.uni-hannover.de}

\begin{abstract}

In recent years, a large number of research efforts aimed at the development of machine learning models to predict complex spatial-temporal mobility patterns and their impact on road traffic and infrastructure.
However, the utility of these models is often diminished due to the lack of accessible user interfaces to view and analyse prediction results.
In this paper, we present the \emph{Traffic Analytics Dashboard (\dashboard)}, an interactive dashboard that enables the visualisation of complex spatial-temporal urban traffic patterns.
We demonstrate the utility of \dashboard at the example of two recently proposed spatial-temporal models for urban traffic and urban road infrastructure analysis.
In particular, the use cases include the analysis, prediction and visualisation of the impact of planned special events on urban road traffic as well as 
the analysis and visualisation of structural dependencies within urban road networks.
The lightweight \dashboard dashboard aims to address non-expert users involved in urban traffic management and mobility service planning.
The \dashboard builds on a flexible layer-based architecture that is easily adaptable to the visualisation of new models.
\end{abstract}

\begin{document}

\maketitle


\section{Introduction}
\label{sec:introduction}

Spatial-temporal traffic analytics can facilitate a variety of applications that require an understanding of traffic patterns in urban regions, including analytics of congestion patterns, planing of road construction sites as well as mobility services and infrastructure planning. 
Development of sophisticated data analytics workflows and specialised machine learning models that facilitate such analytics is an active research area. 
These models utilise heterogeneous data sources, including traffic and mobility data streams, map data (e.g. OpenStreetMap), traffic warnings, accidents, weather data, and event calendars \cite{TempelmeierRLKM19}. 
The results of these models are typically made available in GIS applications as well as in specialised visual analytics tools developed for specific applications \cite{Andrienko17}, which demand significant GIS and data science expertise. 
Recently, efforts has been made to make it easier for non-expert users to create data analytics workflows in the context of mobility analytics \cite{GottschalkTKIFD19}. 
However, for the end-users, including mobility service providers, traffic managers, and city councils, the results of such predictive models are still difficult to access.


We illustrate the problem of spatial-temporal traffic analytics in urban road networks, and in particular analytics of congestion patterns, with two recently proposed analytics models developed in our previous work: 1) prediction of the impact of planned special events on urban road traffic \cite{TempelmeierDD20}, and 2) detection of structured spatial-temporal dependencies in urban road networks \cite{TempelmeierFWD19}.
Planned special events, like concerts, fairs and football games, can significantly affect urban traffic. As observed in \cite{TempelmeierDD20}, such events result in specific congestion patterns that can be learned from historical traffic data effectively. 
Furthermore, as discussed in \cite{TempelmeierFWD19}, urban road networks possess structural dependencies that lead to systematic mutually dependent congestion patterns. 
These models are of particular interest in the context of urban traffic management and planning. Furthermore, these models provide complementary information, potentially leading to a better understanding of congestion patterns in urban areas. 

\begin{figure*}
    \centering
\begin{subfigure}{0.48\textwidth}
    \centering
    \includegraphics[width=\textwidth]{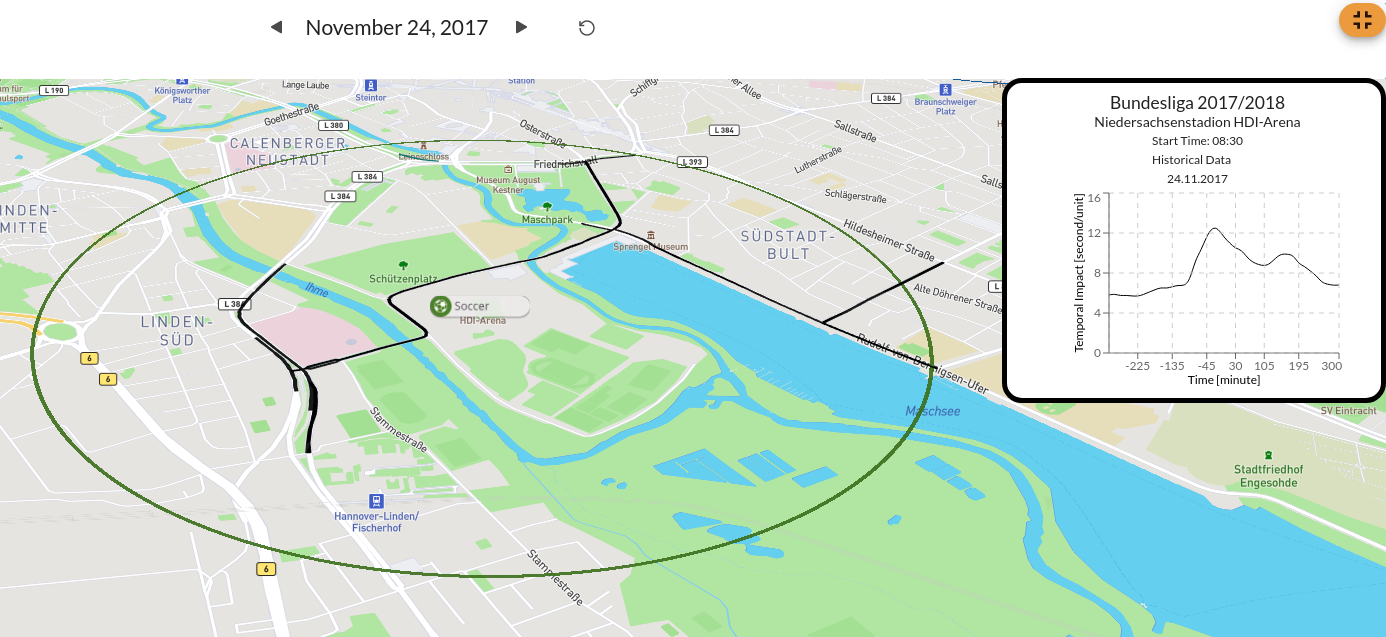}
    \caption{View of historical events with affected subgraph.}
    \label{fig:historical_events}
\end{subfigure}
\begin{subfigure}{0.48\textwidth}
    \centering
    \includegraphics[width=\textwidth]{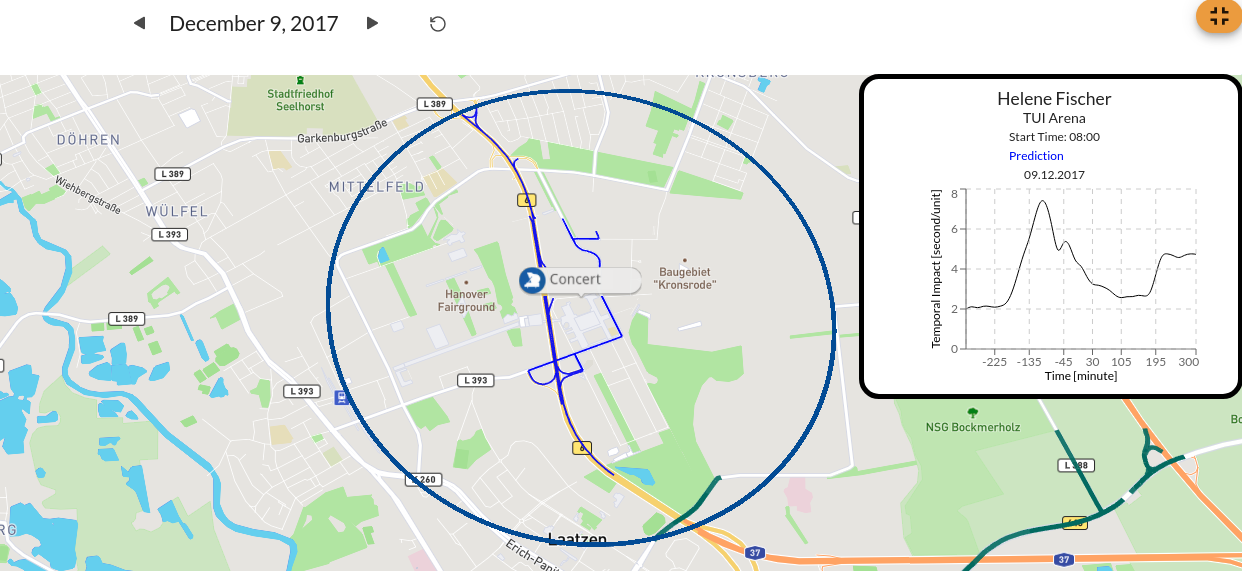}
    \caption{View of future events with typically affected subgraph.}
    \label{fig:future_events}
\end{subfigure}
\caption{View of the event-specific affected subgraph (black graph), typically affected subgraph (blue graph),  spatial event impact (circle), and temporal event impact (plot on the right-hand-side). Map: \copyright OpenStreetMap contributors}
\label{fig:event_view}
\end{figure*}

In this paper, we present \dashboard -- a dashboard that follows a lightweight integration approach to facilitate effective and efficient exploration and joint visual analysis of spatial-temporal traffic patterns discovered by different analytics models on a layered map for end-users. 
Visual analytics has been utilised in the mobility and transportation domain for a variety of purposed, including exploratory analysis of movement data as well as more advanced tasks of traffic modelling, forecasting, and planning \cite{Andrienko17}.
The \dashboard provides an intuitive view on a variety of traffic patterns and does not require any GIS or data science expertise. 
The \dashboard dashboard facilitates lightweight visual analytics on top of heterogeneous urban information sources, including traffic and event datasets, discussed in \cite{TempelmeierRLKM19}.
The current implementation of \dashboard integrates the models for event impact prediction and detection of structured spatial-temporal dependencies. Furthermore, the \dashboard is easily extendable and provides a predefined set of layer templates for the visualisation of additional models. 
Figure \ref{fig:event_view} presents a \dashboard overview, described in Section \ref{sec:demo} in more detail.

Overall, our contributions include: 
\begin{itemize}
    \item We present \dashboard - a lightweight open source dashboard that facilitates joint exploration and visual analytics of patterns resulting from several spatial-temporal traffic analytics models.
    \item We illustrate the functionality of the models for event impact prediction and discovery of structural dependencies in urban road networks as part of the \dashboard dashboard.
    \item We demonstrate the utility of \dashboard on real-world datasets in the urban region of Hannover, Germany. 
\end{itemize}


\section{Traffic Analytics Models}
\label{sec:models}

In this paper, we illustrate \dashboard at the example of two recently proposed traffic analytics models, developed in our previous work. These models include the prediction of the impact of planned special events on urban traffic presented in \cite{TempelmeierDD20}, and detection of structural dependencies in urban road networks presented in \cite{TempelmeierFWD19}.
Whereas these models have been proposed in isolation, 
their joint visual analytics in  \dashboard can reveal 
unseen congestion patterns due to a combination of impact factors that include planned special events and the road network topology.
In this section, we define the basic concepts and describe these models and their integration in  \dashboard in more detail.

\subsection{Formalisation} 
\label{sec:formalisation}
We represent the road network as a directed multi-graph $TG:=(V,U)$, referred to as a \textit{transportation graph}. 
$U$ is a set of edges (i.e. road segments); $V$ is a set of nodes (i.e. junctions).
We refer to an edge of the transportation graph as a \textit{unit} $u\in U$. 
The traffic flow 
observed on a unit at a particular time point is denoted as \textit{unit load}. 
Formally, $ul(u, t)$ is the traffic load on the unit $u$ at the time point $t \in \mathcal{T}$, where $\mathcal{T}$ denotes the set of time points.
We measure the \textbf{unit load} $ul(u, t) \in [0,1]$ as the relative speed reduction at unit $u$ at time point $t$ with
respect to the speed limit $lim(u)$ of the corresponding edge of the transportation graph:
\[ul(u,t)= \frac{lim(u) - speed(u, t)}{lim(u)}.\]

A unit that exhibits an abnormally high traffic load 
at a certain time point $t$ is referred to as an \textit{affected unit} at time point $t$.   
Formally, \au: $U \times \mathcal{T} \mapsto  \{\textit{True}, \textit{False}\}$ indicates whether unit $u$ is affected at time point $t$.

We identify so-called \emph{affected units}, i.e. the units that exhibit an exceptionally high load, by employing the \emph{interquartile range} (IQR) rule  \cite{iqrbook}. 
Weekday and day time strongly influence the traffic load. Therefore, we treat each combination of weekday and day time separately.
More formally, we consider $u$ to be affected at time $t$, if the following condition holds: 
\[
\au =
\begin{cases}
    True, &  if~ul(u, t) > Q_3(u, t) + 1.5 \cdot IQR\\
    False,              & \text{otherwise},
\end{cases} 
\]

\noindent where $Q_n(u, t)$ denotes the $n^{th}$ quartile of the unit load on unit $u$ concerning the weekday and day time and $IQR=Q_3(u, t) - Q_1(u, t)$ denotes the interquartile range. 

We further identify \emph{affected subgraphs} which represent subgraphs of the transportation network $TG$ that indicate a high load and contain affected units. The specific definition of an \emph{affected subgraph}, i.e. under which conditions the units are included in the specific subgraph, depends on the specific traffic analytics model. 

\subsection{Impact of Planned Special Events}
The traffic analytics models considered in this section 
aim to capture the impact of large-scale planned special events, for instance, football games, concerts, fairs, 
on road traffic.
To this extent, we derive the impact of an event from unit load analysis and training supervised machine learning models, 
while taking into account the following spatial-temporal patterns: 1) affected subgraphs, 2) typically affected subgraphs, 3) spatial event impact, and 3) temporal event impact.

\textbf{Affected Subgraphs Identification.}
%
Given an event venue and a specific event, we aim to create an event-specific \emph{affected subgraph} 
that includes the part of the transportation graph affected by the event.
We determine the affected subgraph by considering the following factors:
(i) the geographic proximity between a unit $u$ and the event venue;
(ii) the temporal proximity of the time point $t$ to event start time;
(iii) the affectedness \au of unit $u$ at time $t$;
(iv) the connectivity of unit $u$ with other units that indicate event-induced load in the transportation graph.
We employ a graph-based algorithm to compute the affected subgraph that connects affected units 
located in the spatial-temporal proximity of the event, described in detail in \cite{TempelmeierDD20}.
We integrate the affected subgraphs of the events into \dashboard as a simple layer. 

\textbf{Typically Affected Subgraph Identification.}
Given an event venue, specific subgraphs are commonly affected by planned special events that take place in this venue.
To this extent, we define a \emph{typically affected subgraph} (TAS) to be a venue-specific subgraph of $TG$ that consist of units commonly affected in the presence of planned special events at a particular venue. %
We construct the typically affected subgraph from all units present in a certain percentage of affected subgraphs of a particular venue, determined by a threshold.
We integrate typically affected subgraphs of the event venues into \dashboard as a simple layer. 

\textbf{Spatial Event Impact Prediction.}
We define the \emph{spatial event impact} of an event on the
transportation graph $TG$ at time point $t$ as the maximal distance between the event venue and a unit $u_{d}$ of the affected subgraph at which we observe the load induced by the event at $t$. This definition facilitates statements such as ``\textit{The event that takes place at the concert hall will impact the area within 5 km from the venue}''. This way, the affected area around the venue can be avoided by the traffic not directly involved in the event.
We enable the prediction of the spatial event impact for unseen events using a supervised machine learning model described in detail in \cite{TempelmeierDD20}.
We integrate the visualisation of spatial event impact prediction into \dashboard as a simple layer.

\textbf{Temporal Event Impact Prediction.}
To quantify the \emph{temporal event impact}, we make use of the following
methodology: 
(1) we compute the typically affected subgraph of the event venue, and 
(2) we measure the temporal event impact as the average delay in the presence of events on the units contained in TAS.
We enable the prediction of the temporal event impact for unseen events using a supervised machine learning model described in detail in \cite{TempelmeierDD20}.
We integrate the visualisation of the prediction of temporal event impacts into \dashboard as a simple layer. 

\subsection{Detection of Structural Dependencies}
This model addresses the problem of identification of structurally dependent subgraphs in a road network.
We consider subgraphs to be structurally dependent if: 
(1) The subgraphs are in spatial proximity; 
(2) The subgraphs are typically simultaneously affected by increased traffic load; and %
(3) The road network topology causes the correlation of the traffic load on the subgraphs.

The model consists of the following main steps: 
(i) We identify affected units of the transportation graph using traffic flow data. 
(ii) We identify affected subgraphs of the transportation graph using spatial clustering.
(iii) We identify structural dependencies of the identified subgraphs using mutual information.

\textbf{Affected Unit Identification.}
The goal of this step is to identify affected units, i.e. the units that exhibit an exceptionally high load at any single time point.
Following the definition of an affected unit in Section \ref{sec:formalisation}, we identify the outliers of the unit load using the IQR rule concerning each unit, weekday and day time.

\textbf{Affected Subgraph Identification.}
This step aims to identify connected, disjunctive subgraphs of the transportation graph that indicate unusually high unit load.
We approach this goal by conducting spatial clustering of the affected units of the transportation graph. In this step, the clustering is performed independently at each point in time following the region growing principle \cite{Bins:96}. 
We further define the spatial clusters to affected subgraphs by conducting spatial merging of overlapping subgraphs, employing an incremental greedy approach described in detail in \cite{TempelmeierFWD19}.

\textbf{Structural Dependency Identification.}
In this step, we bring the temporal dimension into consideration and aim to identify the pairs of subgraphs that are typically simultaneously affected.
We compute a subgraph dependency score for each candidate pair.
The intuition behind this score is to capture both the temporal co-occurrence and the spatial proximity of the subgraphs. 
We employ temporal \emph{mutual information} to measure the temporal co-occurrence of the subgraph pairs. 
%
The score is then computed as a combination of the mutual information and an inverse spatial distance metric.
We integrate the visualisation of the structural dependencies into \dashboard as a simple layer. 
\section{Implementation}
\label{sec:dashboard}
The \dashboard builds on a flexible layer-based architecture, easily adaptable to the visualisation of new models.
This section presents the implementation details of \dashboard.
The \dashboard was implemented using a traditional client-server architecture by which the client fetches all relevant data from the backend.
Both backend and client are implemented in ECMAScript (Javascript). 
The client state management was implemented and simplified using redux.
A component-based architecture using react makes up the core of the \dashboard.
The prominent map component uses deck.gl to render individual layers onto the vector-based map provided by mapbox.
The Graphical User Interface (GUI) allows the user to adjust which layers are visible on the map dynamically. 
Each layer contains different model information, model results or datasets.
A lightweight GUI also provides a low entry barrier for potential users.
The models for identification of event impact and structural dependencies are implemented in Java 8. 
The machine learning models for the prediction of event impact were implemented in Python 3.6 using the sklearn library. 
The individual components exchange data in the JSON format.
We make the source code for the dashboard\footnote{\url{https://github.com/Data4UrbanMobility/dashboard}}, the analysis and prediction of event impact\footnote{\url{https://github.com/Data4UrbanMobility/event-impact}} and the structural dependencies\footnote{\url{https://github.com/Data4UrbanMobility/st-discovery}} available under an open source license.



\section{Demonstration Overview}
\label{sec:demo}

In this section, we describe the datasets utilised to train the traffic analytics models presented in Section \ref{sec:models} and the corresponding \dashboard functionalities we will demonstrate.
Our demonstrator is publicly accessible at \url{http://ta-dash.l3s.uni-hannover.de/}.

\subsection{Datasets}
\textbf{Traffic Data.} 
The traffic analytics models presented in this paper have been trained using 
a proprietary traffic dataset that contains aggregated floating car data in the Hannover region, Germany,
available in the context of the Data4UrbanMobility project \cite{Tempelmeier:2019:D4UM}. 
This dataset provides traffic speed records for each unit $u$ of the transportation graph $TG$ in this region.
The dataset covers the period from October 2017 to January 2018 and contains approximately 195 million records in total. 
The records within the dataset contain the average traffic speed on the individual transportation graph units at discrete time points, i.e. $speed(u, t)$, recorded every 15 minutes.

\textbf{Event Data.}
We extracted an event dataset containing information regarding
events that took place in the Hannover region, Germany, from various regional
event-related websites. We selected seven venues with a capacity for at least 1000 participants in the
Hannover region.
For the training of our models, we selected events that took place between October 2017 and January 2018.
As currently planned special events are suspended due to the COVID-19 outbreak, 
we demonstrate the prediction of event impact using events that took place in February 2018.

\subsection{Demonstration Scenario 1: Analysis of Historical Event Impact}
This scenario presents the visualisation of historical events and their respective impact, as illustrated in Figure \ref{fig:historical_events}.
In this scenario, we assume that traffic data is available for historical events, 
such that affected subgraphs and event impact can be determined precisely.
Single event venues and the events that take place at these venues at a specific time are displayed on the map. 
The user can select an event of interest by clicking on the corresponding event venue.
Once an event is selected, the following information is presented to the user: 
1) The event-specific affected subgraph is drawn on the map, with affected units marked in black.
2) The spatial impact of the event is shown as a radius of the circle drawn around the event venue.
3) An infobox on the right side of the interface shows the name, the category, the start time and the temporal event impact. 
The temporal event impact is presented as a line plot. 
The X-axis indicates the time difference to the event start in minutes, where 0 corresponds to the event start time. 
The Y-axis indicates the temporal event impact. 
The user can navigate the days for which the events are displayed by using the arrows located at the top of the user interface.
Figure \ref{fig:historical_events} illustrates the user interface for historical 
events with a single event, a football game, selected.

\subsection{Demonstration Scenario 2: Event Impact Prediction}
This scenario presents the visualisation of the prediction of the event impact for 
future events, as illustrated in Figure \ref{fig:future_events}. 
In this scenario, traffic data from the event time period is not required.
The spatial and temporal event impact are predicted using supervised machine learning models that use historical information regarding event impact for training.
%
Moreover, we display the typically affected subgraph of the specific event venue.
Similar to the historical use case, single events can be selected visually on the map.
After the user selects an event of interest, \dashboard presents the following information:
1) The typically affected subgraph of the event venue is drawn on the map, where the affected units are marked in blue.
2) The predicted spatial event impact is shown as a circle centred at the event venue.
3) The predicted temporal impact is presented as a line plot at the right side of the interface.
Furthermore, the caption ``Prediction'' in the infobox indicates that the information displayed is based on a prediction model.
Figure \ref{fig:future_events} illustrates the user interface for a future event with a single event selected.

\subsection{Demonstration Scenario 3: Structural Dependencies}
\begin{figure}
    \centering
        \begin{tikzpicture}
        \node[inner sep=0pt] (picture) {    \includegraphics[width=0.48\textwidth]{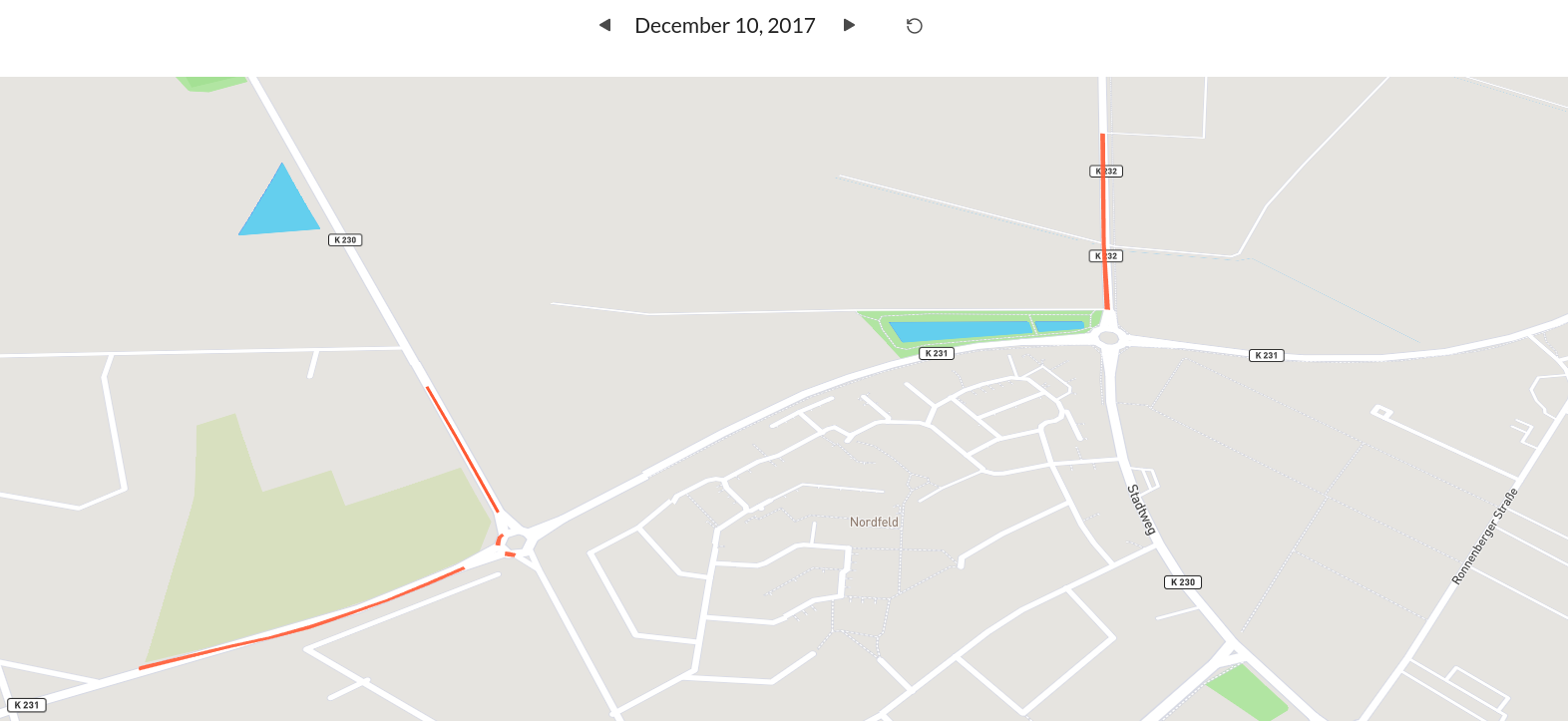}};
        \node[above left, rectangle, fill=white, opacity=0.6, text opacity=1] at (picture.south east) {\tiny \textcopyright OpenStreetMap contributors, ODbL};
    \end{tikzpicture}%

    \caption{View of structural dependencies marked in orange.}
    \label{fig:struct_dep}
\end{figure}

In this scenario, structural dependencies are visualised on the map, as illustrated in Figure \ref{fig:struct_dep}. 
Roads for which such dependencies have been identified are marked in teal colour on the map.
The user can select marked roads with the mouse. 
Once a road is selected, the roads which indicate an interdependence with the selected road are marked in orange.

\section{Conclusions}
\label{sec:conclusion}

In this paper, we presented  \dashboard  - an interactive dashboard that enables the visualisation of complex spatial-temporal urban traffic patterns. 
\dashboard provides a lightweight map-based layered user interface to analyse spatial-temporal traffic patterns in urban areas.
We demonstrated application of \dashboard in several use cases of spatial-temporal traffic analytics in urban road networks, including impact of planned special events and structural dependencies in urban road infrastructure.  
In the future work, we plan to integrate further layers, in particular in the 
context of road safety and electromobility analytics.



\balance

\section*{Acknowledgements}
This work is partially funded by the Federal Ministry of Education and Research (BMBF) and the Federal Ministry for Economic Affairs and Energy (BMWi), Germany under the projects "Data4UrbanMobility" (grant ID 02K15A040), 
"Simple-ML" (grant ID 01IS18054), "CampaNeo" (grant ID 01MD19007B) and "d-E-mand" (grant ID 01ME19009B).

\bibliographystyle{ACM-Reference-Format}
\bibliography{ref}

\end{document}